\documentclass{article}

\usepackage{PRIMEarxiv}

\usepackage[utf8]{inputenc} 
\usepackage[T1]{fontenc}    
\usepackage{hyperref}       
\usepackage{url}            
\usepackage{booktabs}       
\usepackage{amsfonts}       
\usepackage{nicefrac}       
\usepackage{microtype}      
\usepackage{lipsum}
\usepackage{fancyhdr}       
\usepackage{graphicx}       
\graphicspath{{media/}}     

\title{RFUds - A Brain Metastases Imaging Dataset of Radiotherapy Follow-Up}

\author{
    Margarida Fernandes \\
    Center Algoritmi / LASI, University of Minho \\
    Braga, Portugal\\
    \And
    José Soares\\
    Education Manager, Brainlab\\
    Munich, Germany\\
    Postdoctoral Researcher, University of Minho\\
    Braga, Portugal\\
    \And
    Matheus Silva \\
    Director Radiotherapy Applications, Brainlab\\
    Munich, Germany\\
    \And
    Crystian Saraiva\\
    Medical Physicist/Radiotherapy, HCor\\
    São Paulo, Brazil\\
    \And
    Victor Alves \\
    Center Algoritmi / LASI, University of Minho \\
    Braga, Portugal\\
}

\begin{document}
\maketitle

\begin{abstract}
Brain metastases are a common diagnosis that affects between 20\% and 40\% of cancer patients. Subsequent to radiation therapy, patients with brain metastases undergo follow-up sessions during which the response to treatment is monitored. In this study, a dataset of medical images from 44 patients with at least one brain metastasis and different primary tumor locations was collected and processed. Each patient was treated with either a linear accelerator or a gamma knife. Computed Tomography (CT) and Magnetic Resonance Imaging (MRI) scans were collected at various time points, including before treatment and during follow-up sessions. The CT datasets were processed using windowing and artifact reduction techniques, while the MRI datasets were subjected to CLAHE. The NifTI files corresponding to the CT and MRI images were made public available. In order to align the datasets of each patient, a multimodal registration was performed between the CT and MRI datasets, with different software options being tested. The fusion matrices were provided together with the dataset. The aforementioned steps resulted in the creation of an optimized dataset, prepared for use in a range of studies related to brain metastases. RFUds is publicity available at zenodo under the DOI \href{https://zenodo.org/records/14525197}{10.5281/zenodo.14524784}.
\end{abstract}


\newpage
\section{Brain Metastases}
Brain Metastases (BM) represent a frequent occurrence in cancer and constitute the predominant category of tumor in the brain, being five times more common than primary brain tumors \cite{catorze}. It is estimated that approximately 20\% to 40\% of cancer patients experience brain metastases and their incidence has shown a rising trend over time \cite{sete}. In the vast majority of cases, patients develop multiple metastatic tumors in different areas of the brain, known as multiple brain metastases \cite{quinze}.\newline
Metastatic brain cancer occurs when cancer cells move from some primary cancer to the brain. Cells pass through the bloodstream and enter the central nervous system through a breakdown of the blood-brain barrier. This barrier is responsible for protecting the central nervous system and controlling movements between the systemic circulation and neuronal tissues \cite{catorze}, \cite{dezasseis}. Lung, breast and skin primary cancers are most likely to form brain metastases \cite{sete}, \cite{dezassete}. The risk of developing this disease not only depends on the type of primary tumor, but also other factors \cite{dezoito}. Diagnosing this type of cancer involves medical analysis and clinical examinations to explore the presence of abnormal structures inside the brain. The most common diagnostic method is Magnetic Resonance Imaging (MRI), the standard imaging modality for assessing the presence, amount, size and location of BM \cite{dezanove}, \cite{vinte}, usually accompanied by Computed Tomography (CT) acquired with intravenous contrast.\newline
Numerous considerations come into play when evaluating the optimal treatment for brain metastases, including surgery, chemotherapy, targeted therapy, immunotherapy and radiation therapy (Whole Brain Radiation Therapy (WBRT), Stereotatic Radiation Therapy (SRT) and Stereotatic Radiation Surgery (SRS)) \cite{vinte}.

\section{Radiation Therapy}
Radiation therapy, commonly known as radiotherapy, is a method for cancer treatment and it administers beams of ionizing radiation, e.g. x-rays, gamma-rays, electron beams, protons or heavy ions, which damage the DNA of tumor cells, preventing them from dividing or causing them to die \cite{tres}. This treatment has no immediate effect, as it can take a long time for the DNA to change, and it can be combined with other treatments methods. When compared to other treatments, radiotherapy has the huge advantage of being a local treatment, which means it does not affect the whole body \cite{dezoito}. Therefore, radiotherapy is not the best option when the tumor has already spread to various parts of the body, but it is used to shrink early-stage cancer, to stop cancer from recurring, to treat symptoms caused by advanced cancer and to treat cancer that has returned \cite{dezoito}. Depending on the type, size and location of the tumor as well as patient’s general state of health and medical history, there are two main types of radiation therapy, namely external beam radiation therapy and internal beam radiation therapy \cite{tres}.\newline
Internal beam radiation therapy consists of placing a radioactive source inside the body, in an area close to the tumor and it is suitable for small cancers. Depending on whether the source is solid or liquid, the treatment goes by different names, brachytherapy and systemic therapy, respectively. On the other hand, external beam radiation therapy consists of defining the incidence of radiation directed at the tumor, through a machine external to the body and it is often used to treat tumors in the head and neck area, breast, lung, colon and prostate \cite{vinteeum}. The radiation levels employed vary according to the tumor’s specific location. To treat superficial tumors, low-energy radiation is preferable, as it doesn’t penetrate deep into the body. On the other hand, high energy radiation is recommended for treating deeper cancers \cite{vinteedois}.\newline
In order to treat brain metastases, there are different types of radiotherapy treatments that can be applied, namely Whole Brain Radiation Therapy (WBRT), Stereotactic Radiation Therapy (SRT) and Stereotactic Radiation Surgery (SRS). In the first case, radiation is administered to the entire brain, which means that not only tumor cells, but also normal tissue receives radiation. Consequently, it must be divided into several sessions to allow the healthy tissue to recover between each dose \cite{quinze}. However, WBRT reduces neurocognitive function, causing symptoms e.g. fatigue, somnolence and learning and memory difficulties \cite{dezoito}. Regarding SRT and SRS, both options work the same way, applying radiation precisely to the tumor, ensuring that adjacent healthy tissue is not affected or is affected as little as possible. The main difference between these treatments lies in their scheduling, considering that SRT involves a fractionated treatment schedule while in SRS a large dose of radiation is administered in a single session \cite{vinteetres}. There are some different types of machines to administer these treatments, specifically Gamma Knife Systems and Linear Accelerators, being the last one the most commonly employed \cite{vinteequatro}.

\newpage
\subsection{Gamma Knife}
Depending on the model, a Gamma Knife system contain 192 or 201 tiny beams of gamma-rays. In equipment with 201 sources, the beams are emitted from circular pinholes in a fixed helmet worn by the patient. With regard to the equipment with 192 sources, the beams are emitted from circular pinholes fixed in the radiation unit. These beams are focused on the tumor with extreme accuracy, allowing it to receive a high dose of radiation, while the healthy tissue crossed by the beams is subjected to very low doses of radiation \cite{vinteequatro}. Gamma Knife systems emit radiation in the form of ellipsoid (shots). Depending on the shape of the tumor, different shots are needed to mimic its shape. Through technological advancements, it is possible to automatically adjust the width of beams, improving the effectiveness of the treatment \cite{vinteecinco}.\newline
Gamma Knife machines are used to treat small and medium-sized tumors and treatment is done in one day \cite{vinteequatro}. Multiple brain metastases can be treated using this process in a single treatment, with each metastasis being treated individually \cite{vinteecinco}.

\subsection{Linear Accelerator}
Linear Accelerator (LINAC) machines use high energy x-rays to treat tumors. The high energy photons, which are the source of radiation, are generated by the collision of electrons, accelerated by microwave technology, with a heavy metal target located in the LINAC machine \cite{vinteeoito}. A LINAC consists of a system, called gantry, that rotates around the patient, defining irradiation arcs. Considering non-coplanar arcs, it is possible to define ellipsoidal radiation beams that are delivered to the patient via conical collimators. The tumor is positioned at the center where all the central axis of x-rays converge, known as the isocenter. This ensures the tumor receives a high dose of radiation, while surrounding healthy tissues are exposed to a very low dose \cite{vinteenove}. With the evolution of technology, LINACs nowadays use a multi-leaf collimator, which improves treatment by adapting the shape of the beam to match the shape of the tumor from all beams eye view (BEV) \cite{vinteenove}. With these systems, it is possible to improve the conformity and dose fall-off, important parameters for analyzing a radiosurgery plan.

\section{Original Data}
\label{sec:headings}

Data were provided by a single healthcare institution and consist of 44 patients, with various primary tumors and different amounts of brain metastases, whose images were acquired before treatment and during follow-up appointments. Each patient received a minimum of two treatments, either with LINAC or Gamma Knife, in response to the progression of existing metastases or the appearance of new ones. Following these interventions, patients were monitored through regular follow-up appointments, during which new imaging were typically performed to assess treatment outcomes and track possible metastatic progression.\newline
The data set includes information pertaining to both the treatment and follow-up phases. According to the DICOM RT standard, which establishes guidelines for the transmission of medical imaging and treatment data across different systems, a comprehensive account of a radiosurgery treatment requires the inclusion of at least two series of images and the corresponding DICOM RT files \cite{oitentaecinco}.

\begin{figure}[h!]
    \centering
    \includegraphics[width=0.6\linewidth]{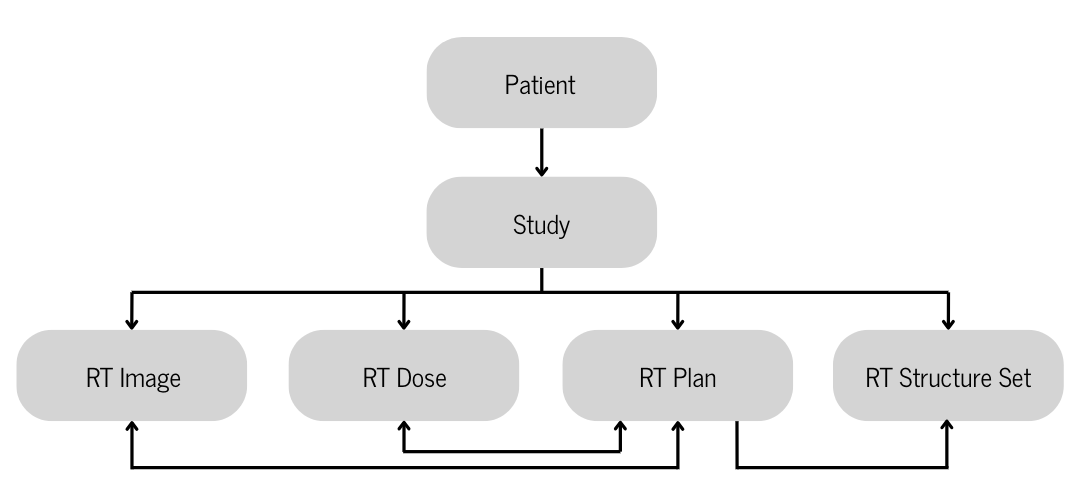}
    \caption{DICOM RT standard objects and respective interconnections. Adapted from \cite{oitentaecinco}.}
    \label{fig:figure_1}
\end{figure}

\begin{figure}[h!]
    \centering
    \includegraphics[width=0.8\linewidth]{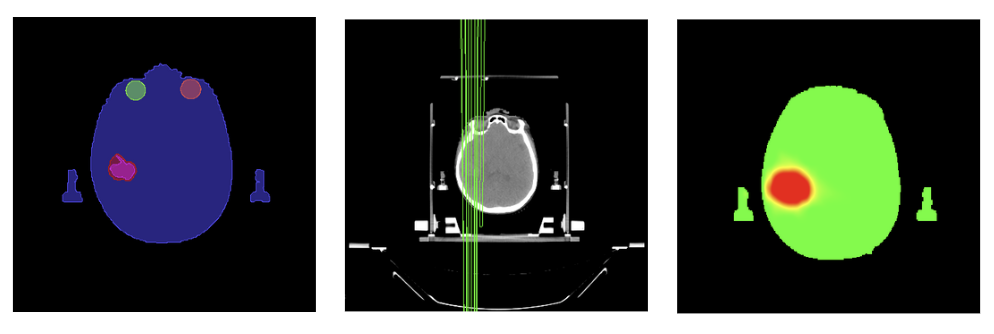}
    \caption{Example of RT Structure, RT Plan and RT Dose files, respectively.}
    \label{fig:figure_2}
\end{figure}

Figure \ref{fig:figure_1} illustrates the DICOM RT objects and the respective interactions. The DICOM RT Dose file provides a comprehensive representation of the dose distribution across the patient’s anatomy. The DICOM RT Structure file contains the delineations of anatomical structures and ROIs that are to be treated, including tumors and OARs. It is usually referenced by the RT Plan file to define the coordinate system and the structures of the patient. Lastly, the DICOM RT Plan file incorporates the treatment plan, including beam parameters, the prescription dose and other details pertaining to the administration of radiation. It is employed during the course of treatment to ensure that the prescribed dose is delivered in accordance with the established plan \cite{oitentaeseis}.\newline
For follow-ups, which occur at varying frequencies depending on the specific needs of the patient, T1 MRI images are available. In most cases, both CT and T1 MRI images are provided for each treatment in DICOM format, along with the corresponding RT Dose, RT Plan and RT Structure files, in DICOM RT format, represented in Figure \ref{fig:figure_2}.

However, for some patients treated with Gamma Knife, CT scans are not included. CT images are typically utilized to provide information regarding electron tissue density information, which is important for the calculation of radiation doses. Nevertheless, Tissue Maximum Ratio (TMR), the algorithm used by Gamma Knife, is able to calculate the dose distribution by assuming that the patient’s head has a uniform electron density, equivalent to that of water. This approach is especially accurate for tumors situated in the central regions of the brain, given that brain tissue is relatively homogeneous. However, it change spread dose for peripheral tumors or in areas with greater heterogeneity, as the presence of air cavities or bones can introduce differences in the dose distribution calculation \cite{oitentaesete}.

\section{Processing Pipeline}
\begin{figure}[h!]
    \centering
    \includegraphics[width=0.6\linewidth]{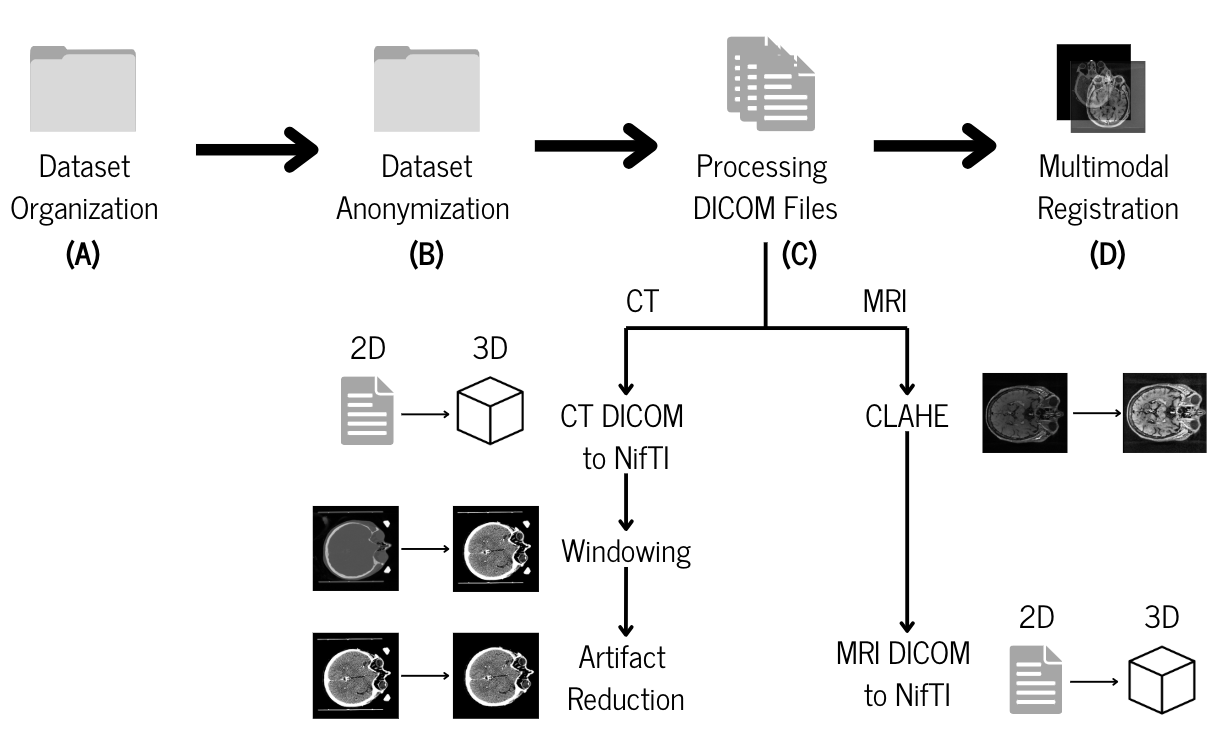}
    \caption{Processing pipeline divided into four steps (A - Dataset Organization, B - Dataset Anonymization, C - Processing DICOM Files, D - Image Registration).}
    \label{fig:figure_3}
\end{figure}

In order to structure the data and prepare it for the next steps, the processing pipeline shown in Figure \ref{fig:figure_3} was followed. Patient data was first organized and properly anonymized. The processing of DICOM files vary between CT and MRI datasets. DICOM CT images are initially converted to NifTI format, then subjected to windowing and subsequently, artifact reduction. DICOM MRI images are processed using CLAHE and then converted to NifTI.

\subsection{Data Organization}
\label{sec:headings}
\begin{figure}[h!]
    \centering
    \includegraphics[width=0.25\linewidth]{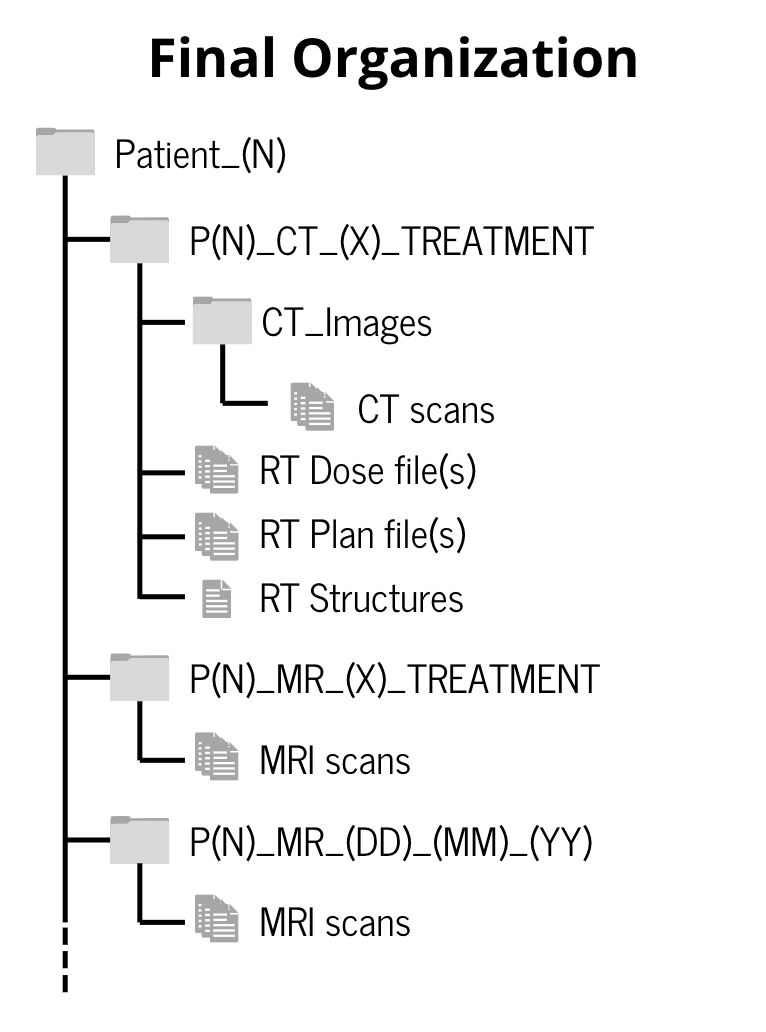}
    \caption{Dataset organization.}
    \label{fig:figure_5}
\end{figure}

To ensure proper organization, a dedicated folder was created for each patient (Patient\textunderscore (N)), containing subfolders that categorize the different types of data. For each treatment, CT images, RT Plan, RT Dose and RT Structure files were stored in one folder (P(N)\textunderscore CT\textunderscore (X) \textunderscore TREATMENT) and MRI images were saved in another folder (P(N)\textunderscore RM\textunderscore (X)\textunderscore TREATMENT). Each follow-up was saved in a folder (P(N)\textunderscore RM\textunderscore (DD)\textunderscore (MM)\textunderscore (YY)) containing the MR images from a specific date (DD-MM-YY). The organization of each patient's data is shown in Figure \ref{fig:figure_5}.

\subsection{Data Anonymization}
\label{sec:headings}
\begin{figure}[h!]
    \centering
    \includegraphics[width=0.35\linewidth]{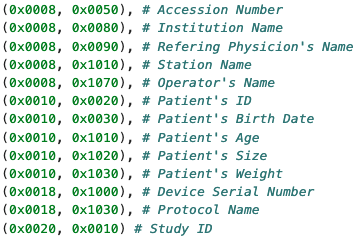}
    \caption{DICOM tags whose values have been deleted.}
    \label{fig:figure_6}
\end{figure}

The DICOM format is the international standard for medical imaging and increases interoperability between devices, simplifying processes involving medical images. DICOM files are created when a healthcare service, e.g. diagnosis or treatment, is performed and consist of two distinct parts, the image pixel data and its associated metadata, the DICOM header \cite{oitentaeoito}.\newline
The header contains specific DICOM elements that complement the image. These elements are composed by a hexadecimal number, the DICOM tag, an optional two-character code, the DICOM Value Representation (VR), the length, which defines the number of bytes of the stored value, and a name \cite{oitentaeoito}. Each DICOM tag represents a specific object that contains information about the institution where the image was acquired, including the institution's name and the operator's name, as well as details about the image acquisition parameters, such as image dimensions and voxel size. Patient health information, e.g. the name, age, sex, medical record and date of birth, is also stored in the image metadata and is used to identify the patient.\newline
In the context of healthcare data management, anonymizing patient data is a crucial step that protects individual privacy and ensures confidentiality (General Data Protection Regulation (GDPR)). The anonymization process, represented by step B in Figure \ref{fig:figure_3}, involved excluding the values associated with certain tags that could identify the patient, from both CT and MRI scans as well as dose, plan and structure files. The tags chosen for removal are presented in Figure \ref{fig:figure_6} \cite{oitentaenove}. The patient’s name, represented by the tag (0010, 0010), was replaced by the patient’s ID and all other applicable tags were left blank.

\subsection{Processing DICOM Files}
\subsubsection{Processing CT Datasets}
\textbf{Converting CT DICOM to NifTI}\newline
The initial step in processing CT files involved converting them to NifTI format. Although DICOM format is highly flexible and complete, it can be complex to implement seamlessly. In comparison, NifTI is a straightforward, simplistic format that has gained wide acceptance in neuroimaging research \cite{noventeeum}.\newline
The first step involved converting the DICOM RT Structure files using the dcmrtstruct2nii function \cite{noventaedois}, that requires as inputs the RT structure file, corresponding image dataset and an output path. This process transforms each outlined structure, e.g. ROI or tumor volume, into a NifTI file, where the structure is assigned a value of 255 (white) and all other areas are labeled 0 (black). Following this conversion, the CT images were also converted using the dcm2niix function. This function produces a NifTI volume representing the CT, along with a Brain Imaging Data Structure (BIDS) JSON file.\newline

\textbf{Windowing Grayscale on CT NifTI}\newline
Hounsfield Units (HU) are a standardized scale used to quantify radiodensity in CT scans. This scale uses the absorption coefficient of radiation in different tissues to create a grayscale image \cite{noventaetres}.\newline
Windowing grayscale is a technique that manipulates the grayscale component of CT images, aiming to emphasize specific structures. Also referred to as gray level mapping, contrast stretching, histogram modification or contrast enhancement, this process improves the brightness and contrast of CT images by manipulating the Window Level (WL) and Window Width (WW), respectively. WW defines the range of CT numbers displayed in the image, whereas WL represents the midpoint of this range. A narrower WW reduces the range of CT numbers, resulting in a quicker transition between dark and light areas. Similarly, a lower WL increases the brightness of the CT image.\newline 
Brain has a window width of 80 HU and a window level of 40 HU \cite{noventaequatro}. Considering these values, the upper and lower gray levels were calculated through equation \ref{eq:equation1} and equation \ref{eq:equation2}, respectively.

\begin{equation} 
\label{eq:equation1}
upper gray level = window level + {\frac{window widht}{2}}
\end{equation}
\begin{equation} 
\label{eq:equation2}
lower gray level = window level - {\frac{window widht}{2}}
\end{equation}

\begin{figure}[h!]
    \centering
    \includegraphics[width=0.8\linewidth]{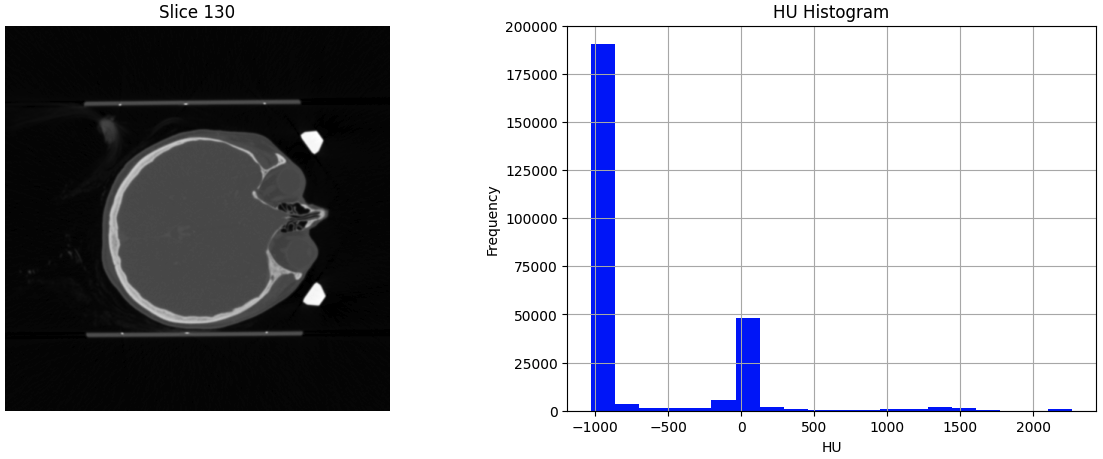}
    \caption{Original CT NifTI image and respective HU histogram.}
    \label{fig:figure_7}
\end{figure}

\begin{figure}[h!]
    \centering
    \includegraphics[width=0.8\linewidth]{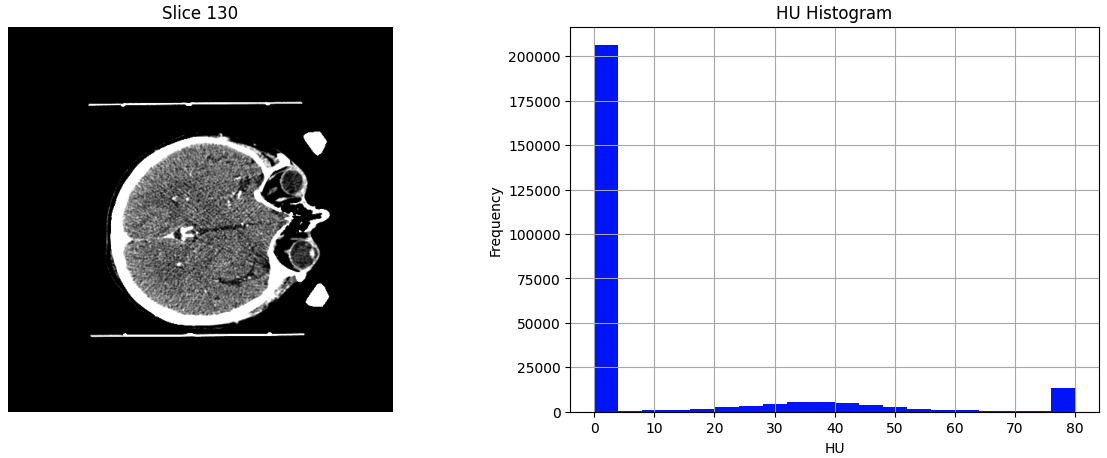}
    \caption{CT NifTI image and respective HU histogram representation after windowing grayscale.}
    \label{fig:figure_8}
\end{figure}

A lower gray level of 0 and an upper gray level of 80 were defined. Voxels within this range were distributed across the full grayscale and voxels above 80 were set to white, the maximum grayscale level, while those below 0 were set to black, the minimum grayscale level.\newline
An example of one slice of a patient is presented before and after windowing, in Figure \ref{fig:figure_7} and Figure \ref{fig:figure_8} respectively, along with the corresponding HU histograms representation. A comparison of the two figures reveals the shift in HU values and its impact on the contrast and brightness of the CT dataset.\newline\newline

\begin{figure}[h!]
    \centering
    \includegraphics[width=0.6\linewidth]{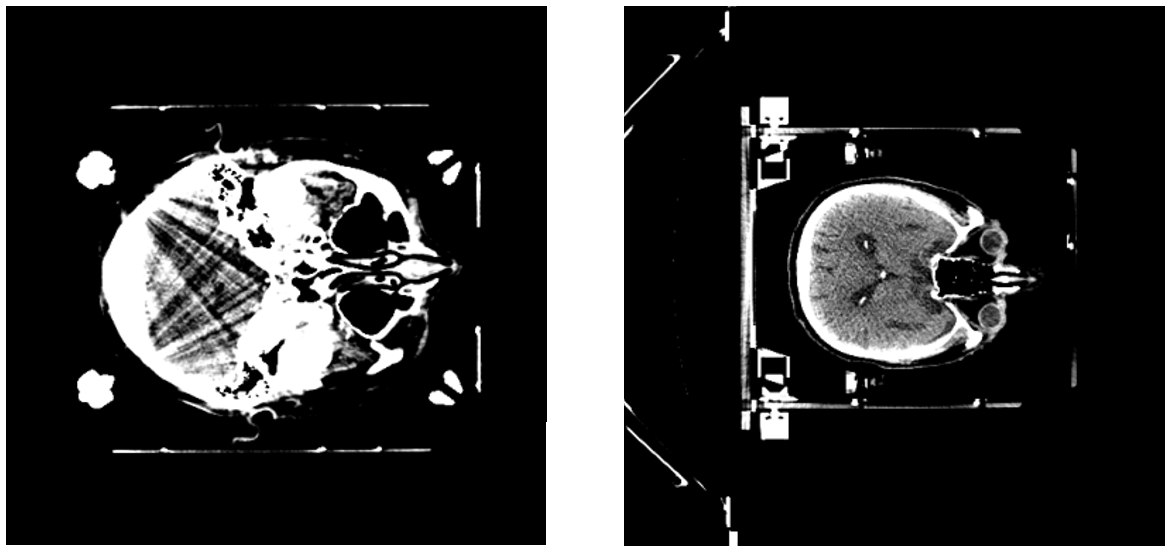}
    \caption{Immobilization for patients treated with Gamma Knife Perfexion (on the left) and for patients treated with LINAC (on the right).}
    \label{fig:figure_9}
\end{figure}

\textbf{Artifact Reduction on CT NifTI}\newline
Frames or thermiplastic masks (frameless) systems are employed to ensure that patients are optimally immobilized during radiotherapy treatment, a critical requirement for an effective procedure. However, while successfully preventing movement, these masks have the disadvantage of introducing some artifacts into CT images. The configuration of these masks can vary depending on the machine used for treatment, as illustrated in Figure \ref{fig:figure_9}.

\begin{figure}[h!]
    \centering
    \includegraphics[width=0.8\linewidth]{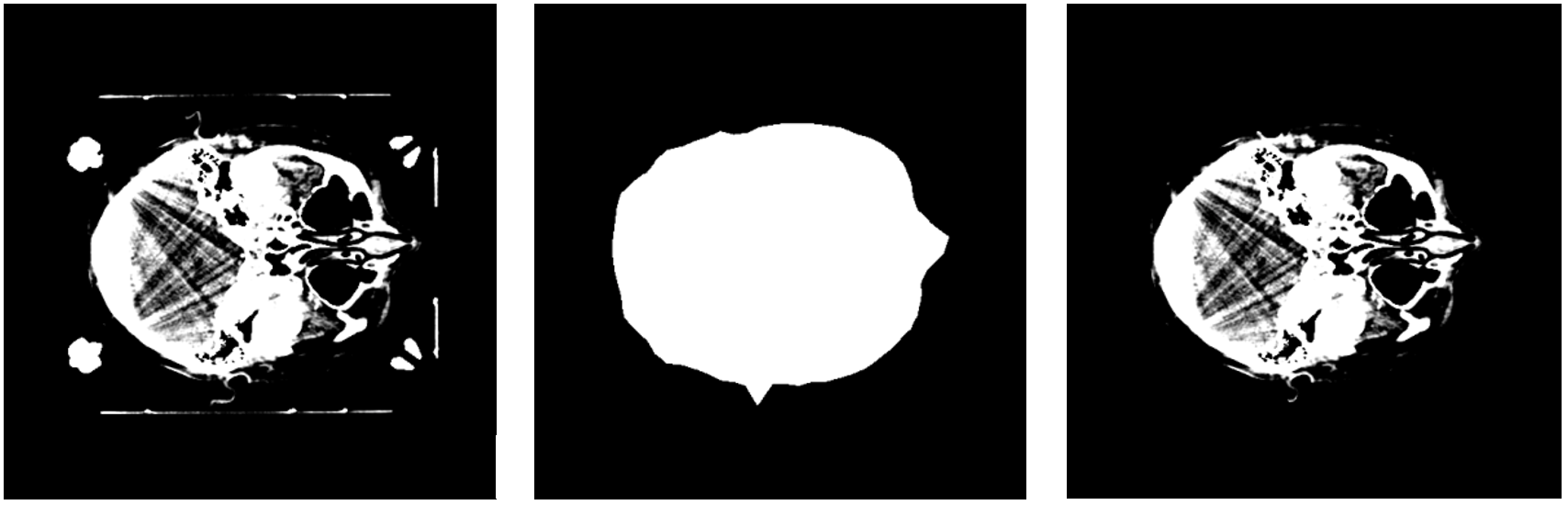}
    \caption{Original image of a patient treated with Gamma Knife Perfexion (on the left) with the respective reflected skull mask (in the middle) and the final image, after artifact removal (on the right).}
    \label{fig:figure_10}
\end{figure}

Considering that artifacts can negatively affect the study of the datasets, their removal was necessary. Depending on whether the patient was treated with Gamma Knife or LINAC, either Skull or Outer Contour masks were applied, respectively. First, the masks were reflected, to align with the direction of the CT dataset, and then acted as filters, allowing only the relevant voxel values to pass unchanged in HU, while rendering the remaining parts of the image in black. Figure \ref{fig:figure_10} shows an example of the original image, the corresponding skull mask and the final image, after artifact removal for a patient treated with Gamma Knife.

\begin{figure}[h!]
    \centering
    \includegraphics[width=0.8\linewidth]{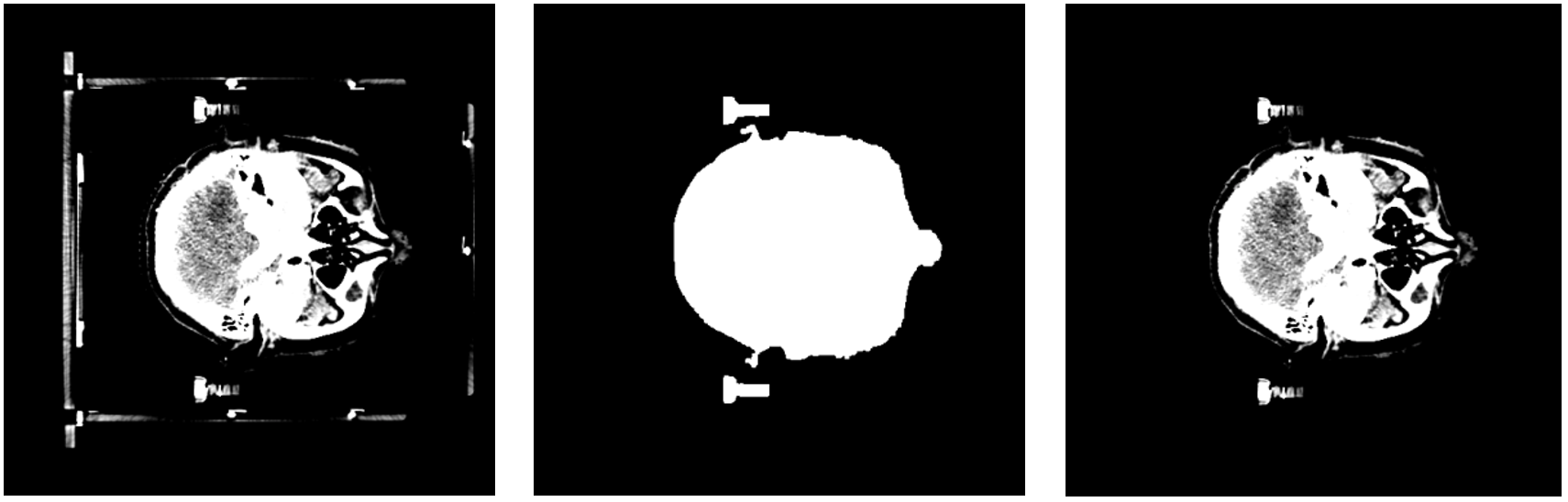}
    \caption{Example of a case where the mask capture regions of the patient’s mask. Original image (on the left) with the respective reflected skull mask (in the middle) and the final image, after artifact removal (on the right).}
    \label{fig:figure_11}
\end{figure}

It is noteworthy that, while the majority of cases demonstrated optimal performance, in some patients, e.g. the one shown in Figure \ref{fig:figure_11}, the mask captured regions of the patient’s immobilization mask, resulting in the retention of these regions in the final image.

\subsubsection{Processing MRI Datasets}
\begin{figure}[h!]
    \centering
    \includegraphics[width=0.8\linewidth]{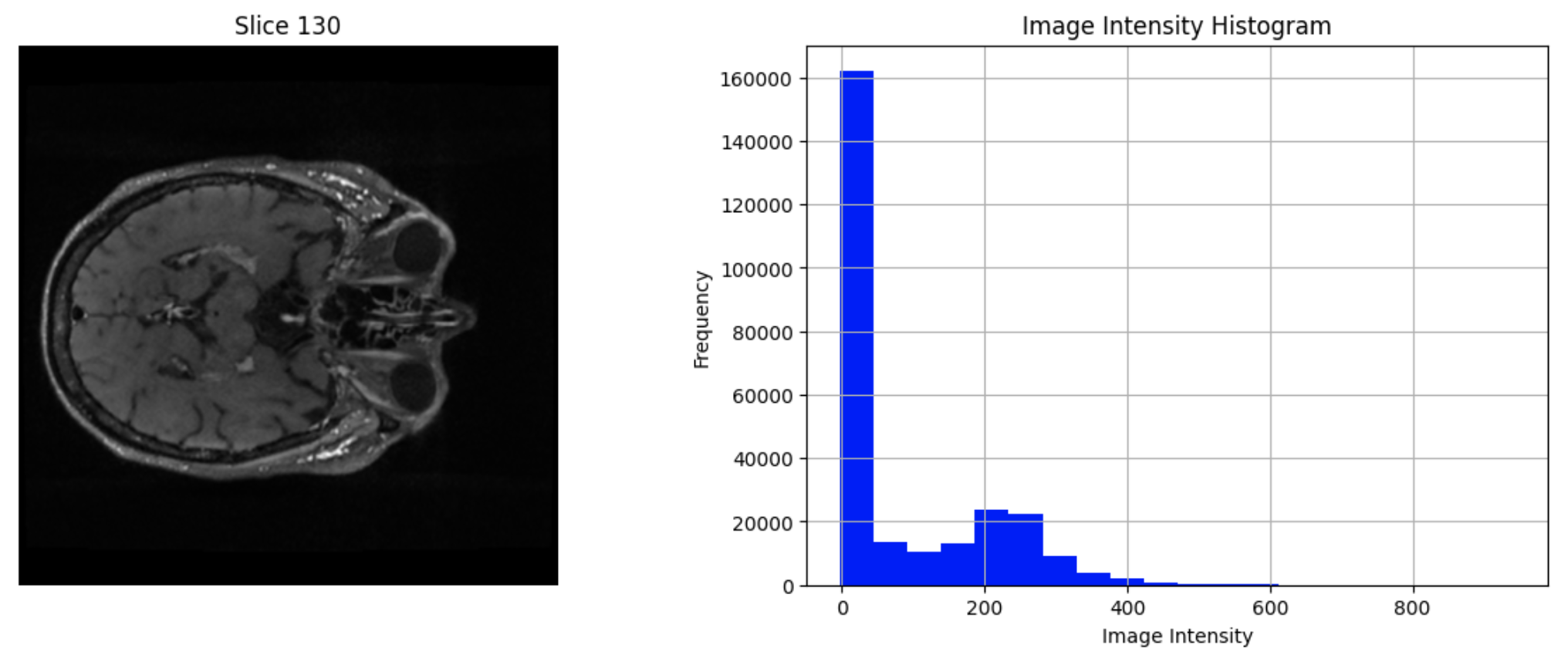}
    \caption{Original MRI image and respective image intensity histogram.}
    \label{fig:figure_12}
\end{figure}

\textbf{Contrast Limited Adaptive Histogram Equalization on DICOM MRI and conversion to NifTI}\newline
Contrast Limited Adaptive Histogram Equalization (CLAHE) is a computational technique used to improve local image contrast. It is based on Adaptive Histogram Equalization (AHE), a process that aims to enhance contrast by dividing the image into small regions and performing histogram equalization of each region. However, this technique introduces noise, a problem solved with CLAHE. The core concept of CLAHE involves applying histogram equalization to independent regions of the image, followed by interpolation to smooth out any inconsistencies along the boundaries between these regions \cite{noventaecinco}, \cite{noventaeseis}. CLAHE is commonly used in medical imaging \cite{noventaeseis} and was used in this study to process MRI DICOM files. Figure \ref{fig:figure_12} shows an example of an MRI image before processing, along with the corresponding intensity values histogram.

\begin{figure}[h!]
    \centering
    \includegraphics[width=0.8\linewidth]{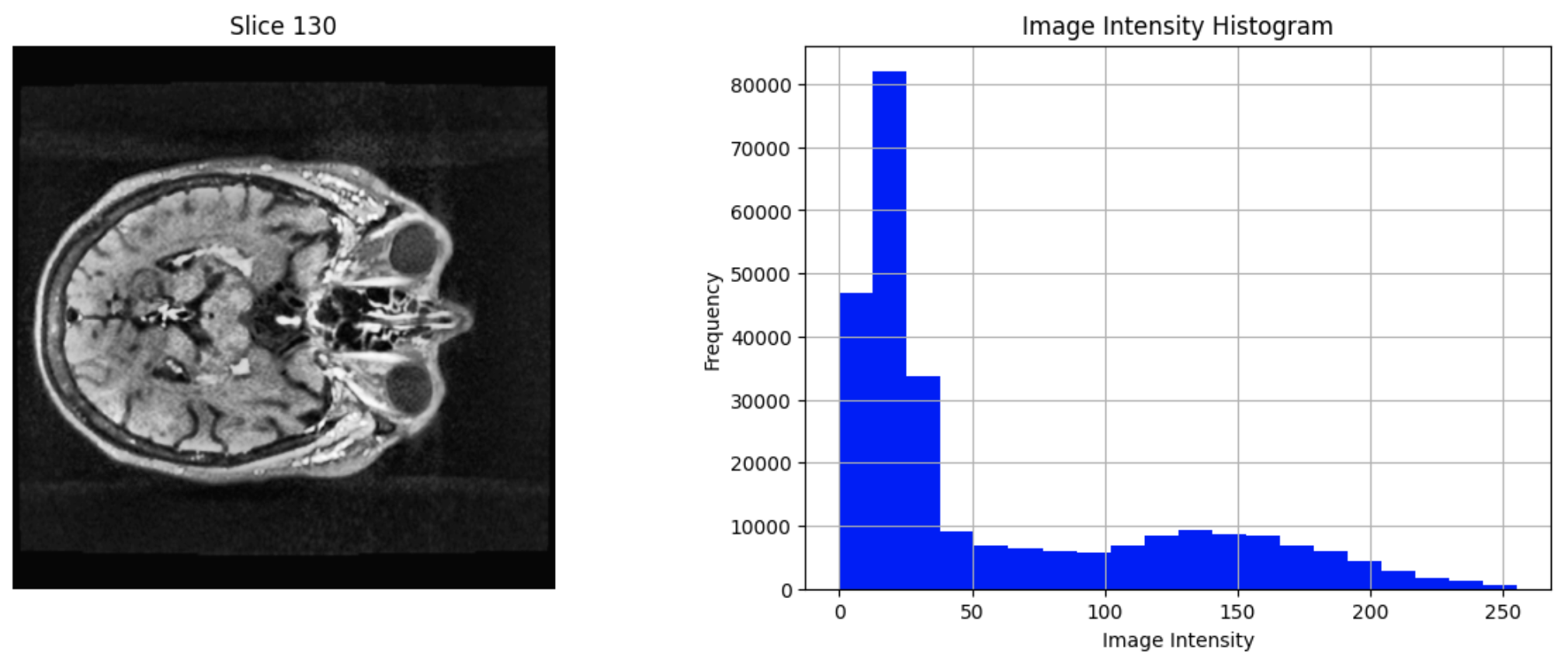}
    \caption{MRI image after processing using CLAHE and corresponding image intensity histogram.}
    \label{fig:figure_13}
\end{figure}

Firstly, the pixel data were normalized to a range of 0 to 255 in order to adjust the intensity and ensure uniform brightness in all images. Then, CLAHE was performed with a clipLimit of 5 \cite{noventa}. This process was applied to each DICOM file using createCLAHE function and the respective headers were updated. The modified images were saved and subsequently converted to NifTI files using the same process as for the CT datasets. Figure \ref{fig:figure_13} represents the previous example post processing and the respective image intensity histogram.

\subsection{Multimodal Registration Process}
\begin{figure}[h!]
    \centering
    \includegraphics[width=1\linewidth]{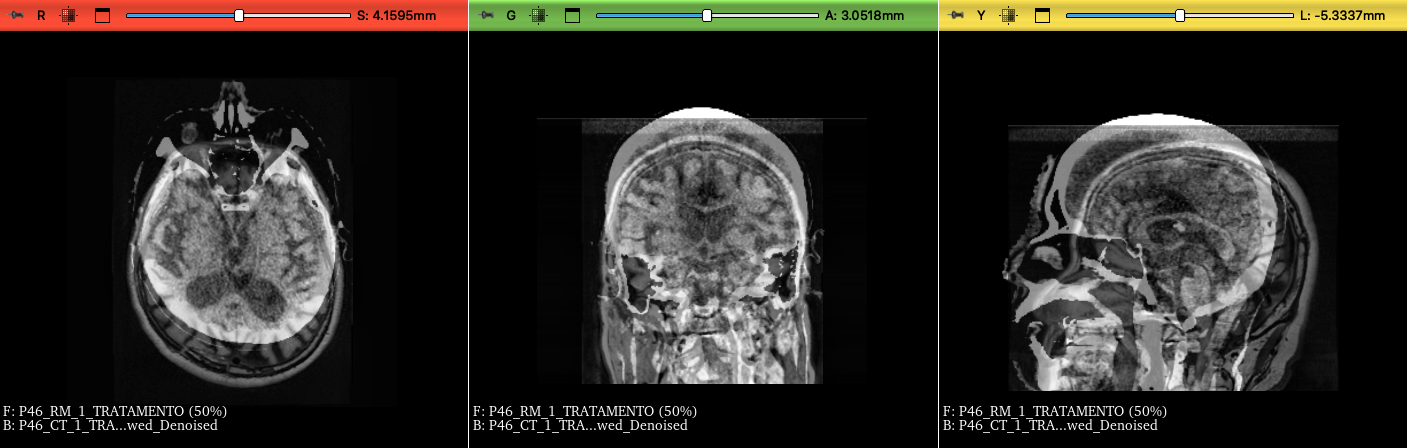}
    \caption{Example of original image set up between CT and MRI datasets.}
    \label{fig:figure_14}
\end{figure}

In order to facilitate a comparison of the images of each patient over time, it is essential to ensure that all datasets from each patient are aligned within the same spatial coordinates, enabling meaningful comparison. However, the datasets from the original data were acquired in different coordinate spaces, making image registration an essential step in the process.\newline 
Image registration is a computational method designed to align images by establishing spatial correspondences between them, ensuring that corresponding points in each image align accurately. The process involves transforming a moving image to spatially match a fixed target image. The alignment is guided by a transformation model and a coefficient, which evaluates the quality of the alignment. The algorithm operates iteratively, adjusting the transformation to minimize the cost function until optimal alignment is achieved \cite{noventaesete}, \cite{noventaeoito}. Figure \ref{fig:figure_14} illustrates an example of the misalignment between the CT and MRI datasets.

In the treatment planning process, MRI datasets are used to outline the organs at risk and the metastases to be treated. On the other hand, CT datasets are utilized for calculating the dose distribution and define stereotaxis for spatial orientation of points, in order to deliver, with precision, the maximum amount of radiation to the tumors while minimizing exposure to healthy tissues. Although structures are delineated on MRI datasets, they are subsequently transferred into the CT coordinate space to enable dose calculation. As a result, the NifTI volumes of the structures outlined during treatment planning are aligned with the CT dataset. For this reason, the MRI datasets were moved for the CT spacial coordinate space. Each MRI dataset was aligned with the previous CT dataset: MRI from treatments were aligned with the respective CT and for follow-ups, the MRI was aligned with the CT from the preceding treatment.\newline
The image registration process, represented by step D in Figure \ref{fig:figure_3}, was evaluated using CT and MRI datasets from the initial treatment of 6 randomly selected patients. This dataset, formally designated as D\textunderscore registration, was employed to test the efficacy of various tools and identify the most reliable option for this crucial step.\newline 
Ensuring accuracy at this stage is essential for maintaining credibility in the subsequent phases of the workflow. FSL, 3D Slicer, ANTs and Brainlab Elements were tested and compared, consistently using 12 degrees of freedom (DOF) for the affine registration. This approach accounts for variations in shape, size and orientation by allowing for translations, rotations, scaling and shearing along all three axes \cite{noventaenove}.

\subsubsection{Image Registration with FSL}
\begin{figure}[h!]
    \centering
    \includegraphics[width=0.8\linewidth]{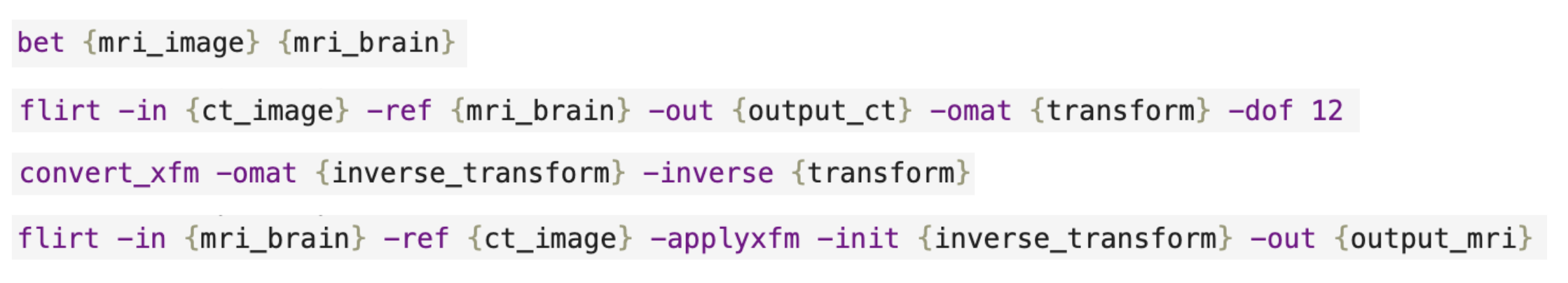}
    \caption{Sequence of commands used for image registration with FSL.}
    \label{fig:figure_15}
\end{figure}

FSL (FMRIB Software Library) is a complete library designed for analyzing functional, structural and diffusion MRI brain imaging data. FMRIB's Linear Image Registration Tool (FLIRT\footnote{More about FLIRT on \href{https://fsl.fmrib.ox.ac.uk/fsl/docs/\#/registration/flirt/index}{FLIRT Documentation}.}) is a fully automated tool from FSL for performing linear image registration \cite{cem}, \cite{centoeum}. Its main options include an input volume (-in) and a reference volume (-ref), which can be from the same imaging modality or from different modalities. The resulting affine transformation that aligns the input with the reference can be saved as an affine matrix (-omat) and an output volume (-out), resulting from applying the transformation to the input image, can be generated. Additionally, parameters e.g. degrees of freedom (DOF) can also be adjusted as needed.\newline 
In a first approach, only the FLIRT command was used to register the MRI and CT datasets of the patients. This tool recommends using the best quality dataset as a reference, so the MRI dataset was used as the fixed image and the CT dataset as the moving image. Since the initial aim was to move the MRI to the CT coordinate space, the inverse transformation matrix was calculated, using convert\textunderscore xfm, and then was applied to the MRI dataset, using applyxfm function.\newline
However, the results of this initial approach were not satisfactory for all patients. Therefore, Brain Extraction Tool  (BET\footnote{More about BET on \href{https://fsl.fmrib.ox.ac.uk/fsl/docs/\#/structural/bet}{BET Documentation}.}) was used prior to executing the previously explained commands. BET separates brain tissue from non-brain tissue \cite{centoetres}. This command was applied to the MRI dataset, and the output served as the fixed reference image for the following steps. Figure \ref{fig:figure_15} illustrates the final sequence of commands for implementing image registration with FSL.

\begin{figure}[h!]
    \centering
    \includegraphics[width=1\linewidth]{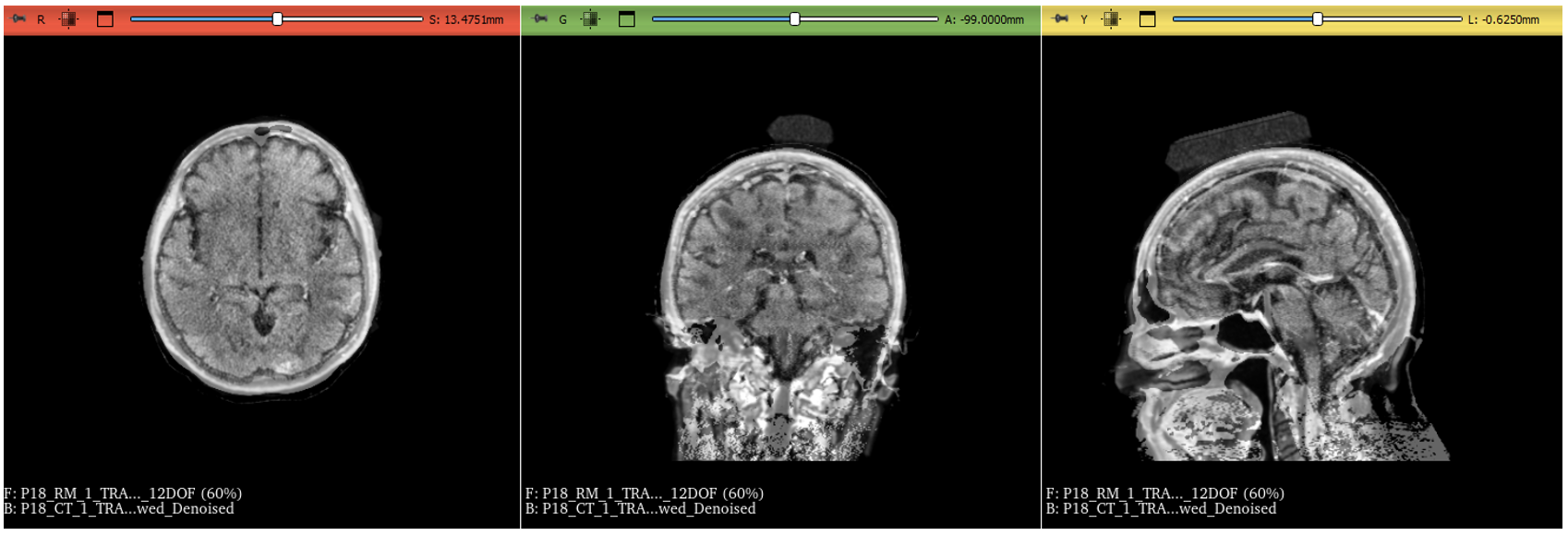}
    \caption{Example of a successful case of image registration using FSL.}
    \label{fig:figure_16}
\end{figure}

\begin{figure}[h!]
    \centering
    \includegraphics[width=1\linewidth]{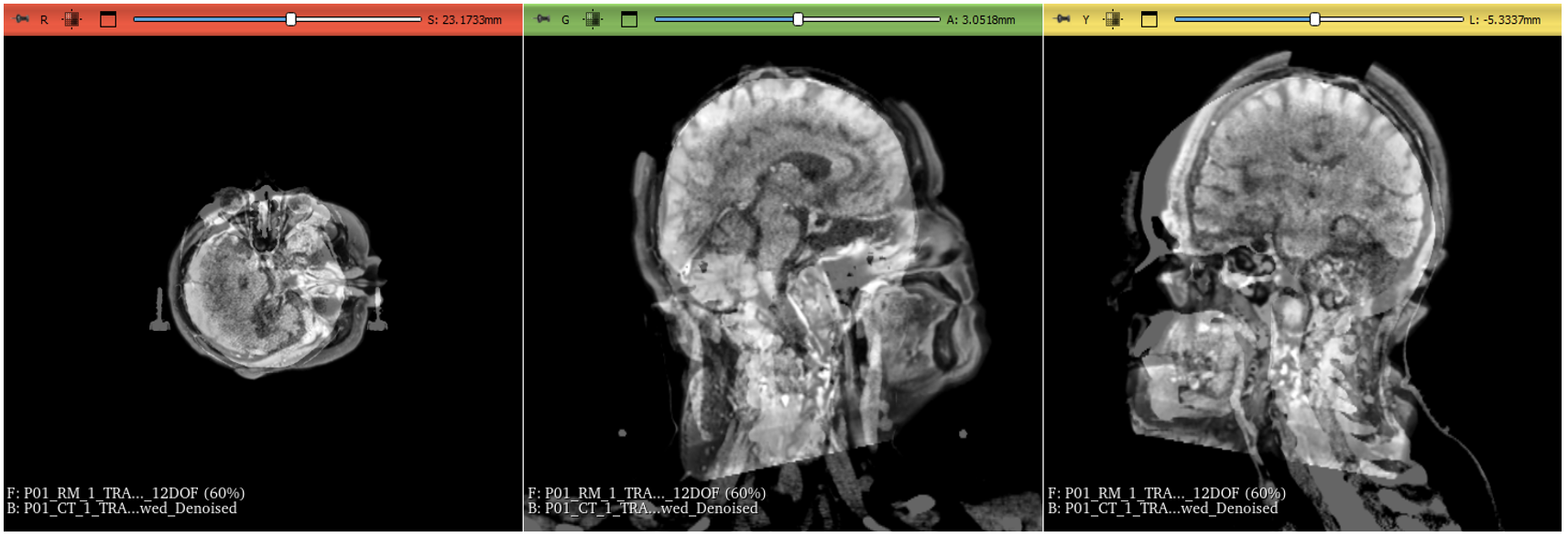}
    \caption{Example of an unsuccessful case of image registration using FSL.}
    \label{fig:figure_17}
\end{figure}

Although the use of BET improved the quality of the registrations, the process became time-consuming and the overall results were unsatisfactory, as the process was only successful in 1 out of the 6 patients from D\textunderscore registration. Figure \ref{fig:figure_16} represents the successful case while Figure \ref{fig:figure_17} exemplifies one of the unsuccessful cases.

\subsubsection{Image Registration with 3D Slicer}
\begin{figure}[h!]
    \centering
    \includegraphics[width=1\linewidth]{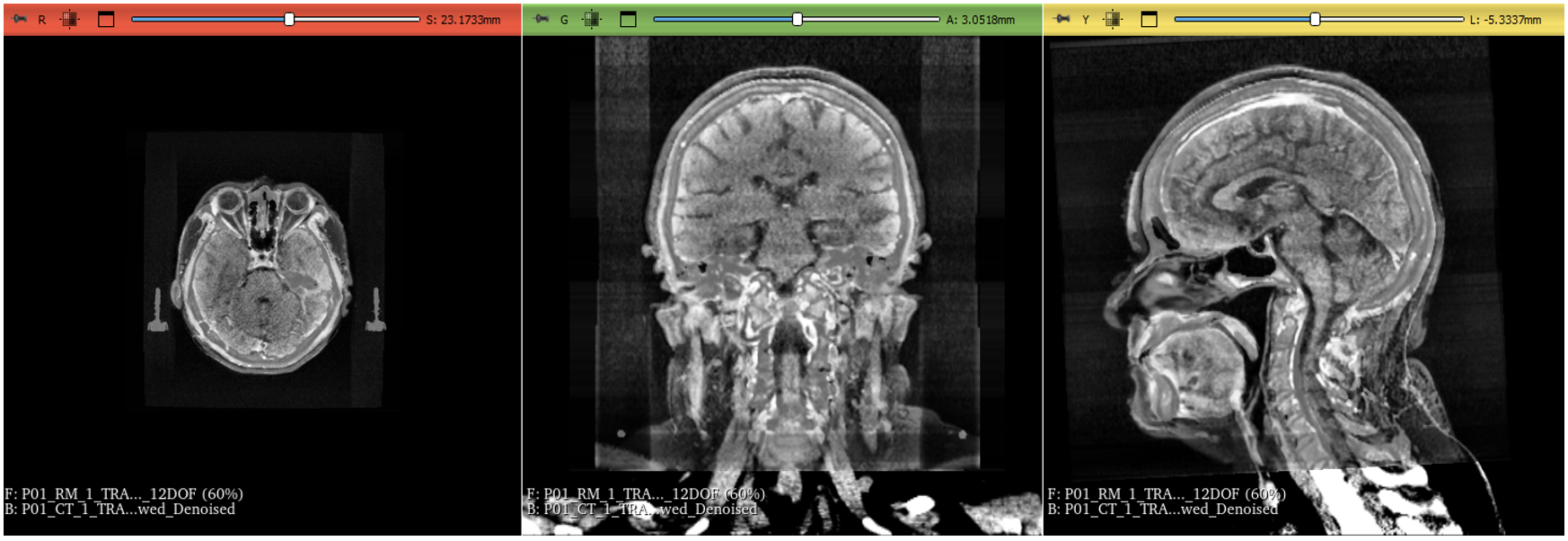}
    \caption{Example of a successful case of image registration using 3D Slicer.}
    \label{fig:figure_18}
\end{figure}

General Registration (BRAINS)\footnote{More about General Registration on \href{https://slicer.readthedocs.io/en/latest/user_guide/modules/brainsfit.html}{Slicer Documentation}.} is a 3D Slicer  module developed to perform three-dimensional image registration between two different datasets. This process requires the selection of two input datasets: a fixed image and a moving image. Additionally, the DOFs and the method for initializing the transform center can be specified. The latter can be selected from a set of available options, that vary according to the respective assumption:
\begin{itemize}
\item useMomentsAlign: the center of mass of the two datasets corresponds to similar structures.
\item useCenterOfHeadAlign: the center of mass is estimated using the top of the head and the shape of the neck.
\item useGeometryAlign: the center of the voxel lattice in the images represents similar structures.
\end{itemize}
Considering the results of the previous attempt and since it is recommended to use the dataset with the most detail as the fixed image, the MRI datasets were used as the reference. The initialize transformation mode was set to useCenterOfHeadAlign and 12 DOFs were selected. After completing this process, the inverse matrices were directly calculated using 3D Slicer and applied to the MRI datasets. Figure \ref{fig:figure_18} represents the only successful case of this method.

\begin{figure}[h!]
    \centering
    \includegraphics[width=1\linewidth]{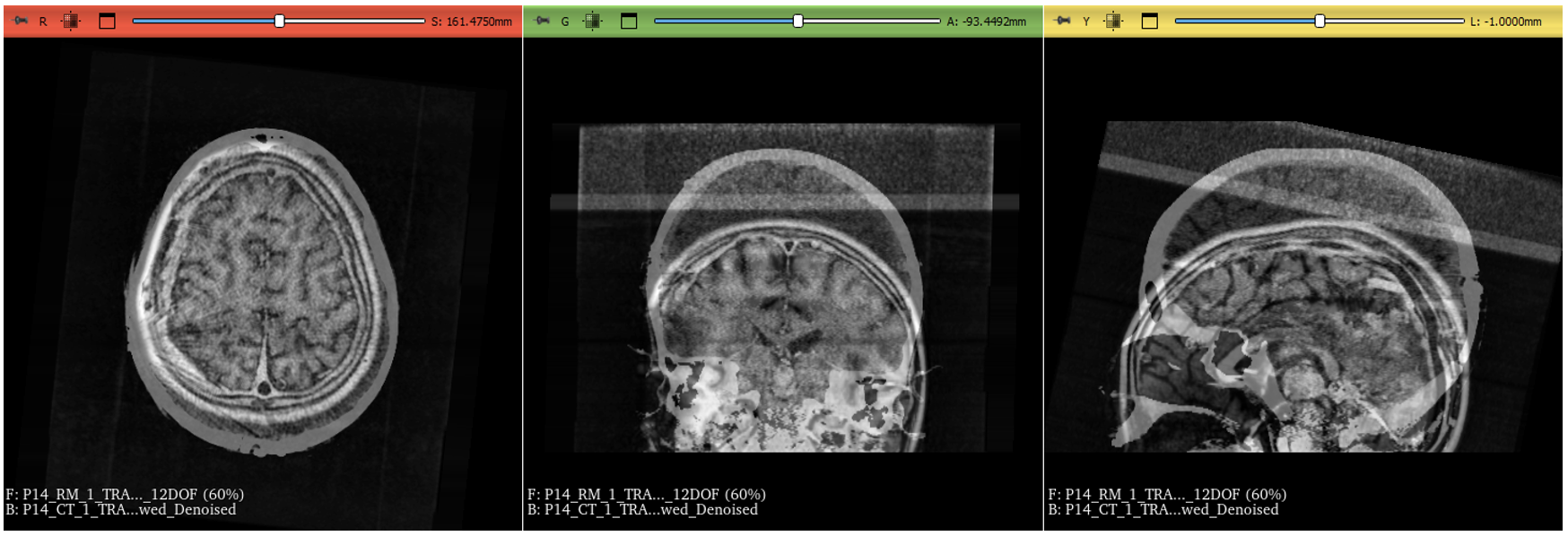}
    \caption{Example of an unsuccessful case of image registration using 3D Slicer.}
    \label{fig:figure_19}
\end{figure}

As the previous method, General Registration (BRAINS) also presented several ineffective cases, represented by the example in Figure \ref{fig:figure_19}.

\subsubsection{Image Registration with ANTs}
\begin{figure}[h!]
    \centering
    \includegraphics[width=0.9\linewidth]{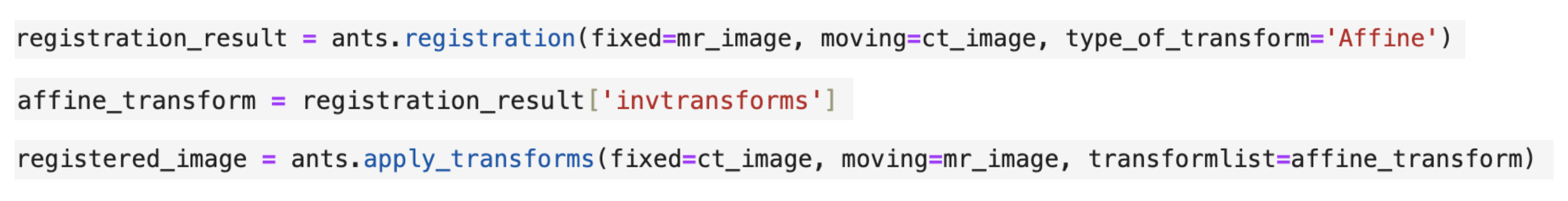}
    \caption{Sequence of commands used for image registration with ANTs.}
    \label{fig:figure_20}
\end{figure}

Advanced Normalization Tools (ANTs)\footnote{More about ANTs on \href{https://github.com/antsx/ants}{ANTs Documentation}.} is an innovative library offering a suite of medical image registration and segmentation tools. It enables users to organize, visualize and analyze statistically large biomedical imaging datasets. Registration  is the image registration function from ANTs that admits a large number of inputs. The fixed volume, the moving volume and the type of transformation are mandatory inputs. Once again, the CT dataset was chosen as the moving image and the MRI dataset as the fixed one and the transformation type was set to affine. The apply\textunderscore transforms function was then used to apply the inverse transformation matrix to the MRI dataset. Figure \ref{fig:figure_20} illustrates the commands used for performing image registration with ANTs.

\begin{figure}[h!]
    \centering
    \includegraphics[width=1\linewidth]{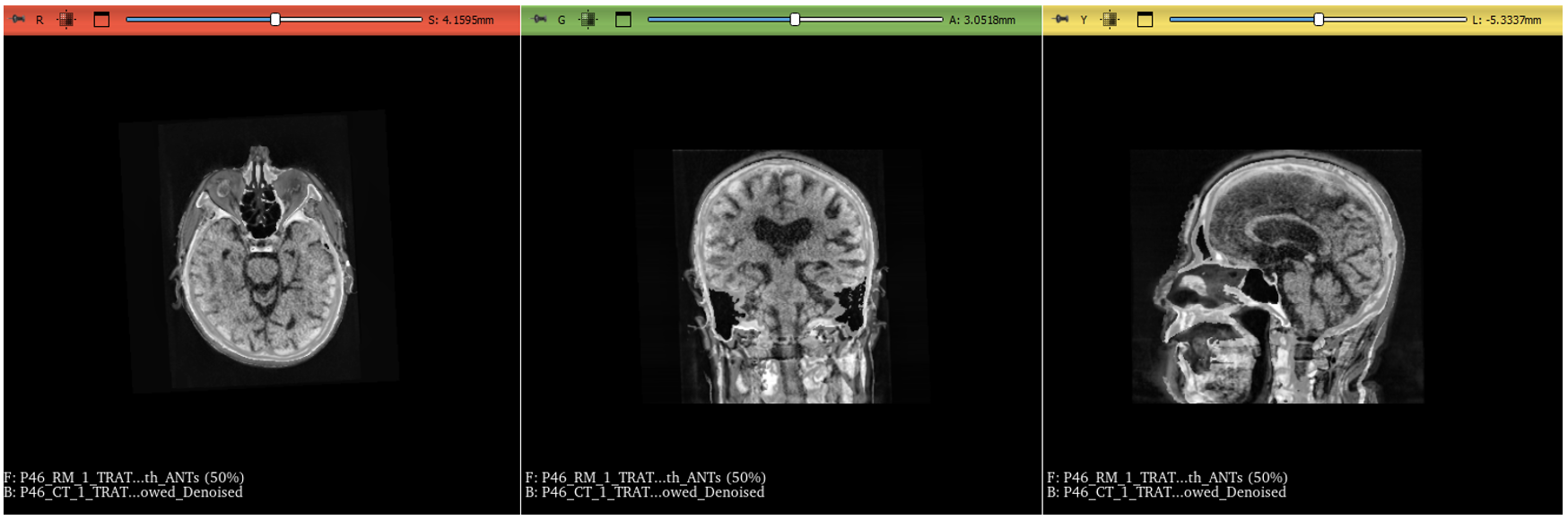}
    \caption{Example of a successful case of image registration using ANTs.}
    \label{fig:figure_21}
\end{figure}

\begin{figure}[h!]
    \centering
    \includegraphics[width=1\linewidth]{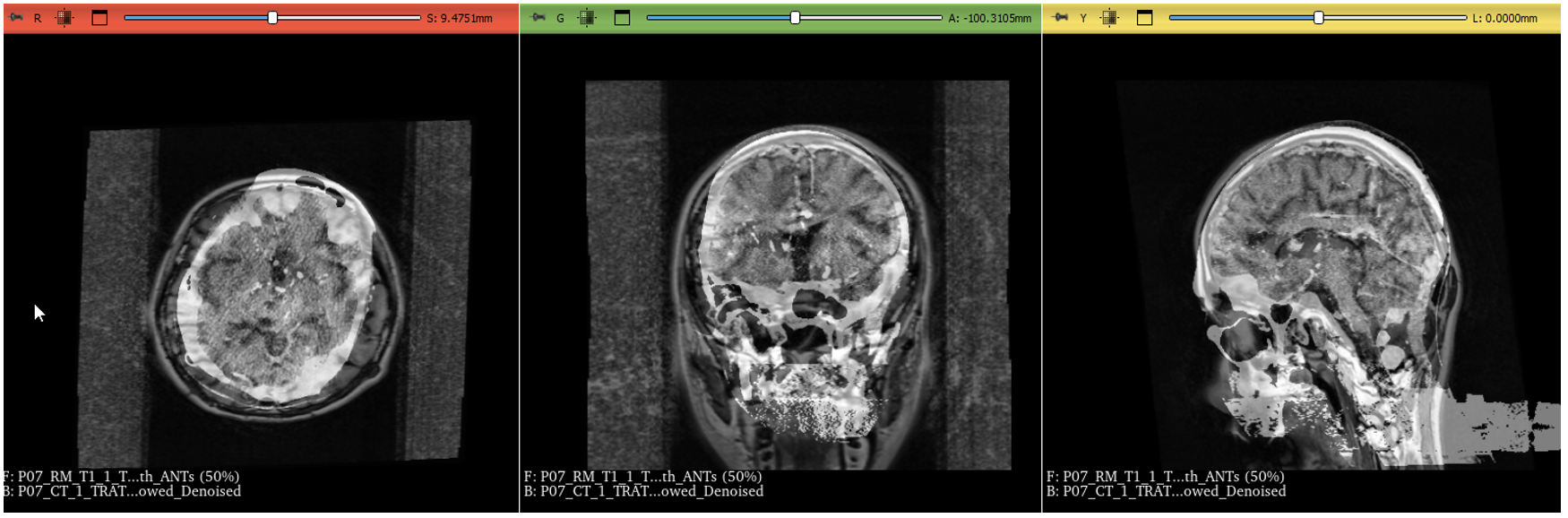}
    \caption{Example of an unsuccessful case of image registration using ANTs.}
    \label{fig:figure_22}
\end{figure}

As in the two previous cases, this method worked in some of the patients from D\textunderscore registration, e.g. the patient shown in Figure \ref{fig:figure_21}, but was ineffective in others, as shown in Figure \ref{fig:figure_22}.

\subsubsection{Image Registration with Brainlab Elements}
\begin{figure}[h!]
    \centering
    \includegraphics[width=1\linewidth]{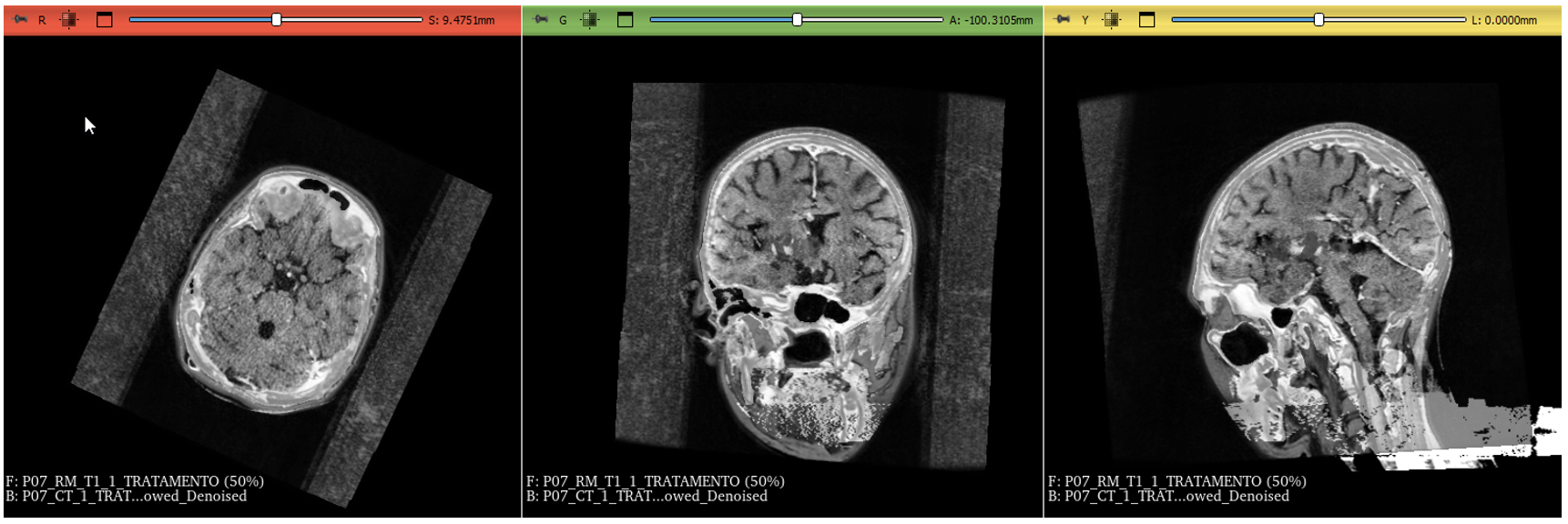}
    \caption{Example of a successful case of image registration using Elements.}
    \label{fig:figure_23}
\end{figure}

Given the not satisfactory results obtained from FSL, 3D Slicer and ANTs, Brainlab Elements was used. Among all the tested options, Elements was the only one that consistently guaranteed a reasonable image registration process for D\textunderscore registration patients. Elements was therefore used to perform image registration across the entire dataset.\newline 
Each MRI dataset was aligned with the corresponding prior CT. Since MRI datasets generally offer more quality than CT datasets, Elements treated the MRI as the fixed image and the CT as the moving image, generating the correspondent transformation matrix. The inverse of each matrix was then calculated in order to apply to the respective MRI dataset. The final transformation matrices for each registration are provided along with the dataset. Figure \ref{fig:figure_23} demonstrate an example of the precise alignment between the CT and MRI datasets after image registration with Elements, in one of the patients whose results with both FSL, 3D Slicer and ANTs were unsatisfactory.

\section{Optimized Brain Metastases Imaging Dataset}
The aforementioned processing pipeline generates a dataset, denoted as D\textunderscore optimized, containing the original images, after processing, in NifTI format, along with volumes concerning the structures outlined during treatment planning. The resulting imaging datasets exhibit enhanced contrast compared to the original DICOM datasets and the CT datasets are devoid of artifacts. In addition, a document containing the transformation matrices resulting from the image registration process is included in the dataset.

\section{Conclusion}
Early detection of brain metastases, followed by a timely intervention, plays a crucial role in mitigating the impact of this condition and significantly increasing the chances of treatment success. In this study, a dataset provided by a single institution was optimized. The flowchart presented for processing the information was adapted so that the information about the images, as well as the dose, plan and structure files, did not lose data. The process of anonymizing the data used in this study made it possible to safely keep inaccessible any data that could identify the patients whose images were used. The data processing stage made it possible to convert DICOM images to NIFIT without loss of information. Furthermore, image quality was improved using windowing on CT datasets and CLAHE in MRI datasets. It was possible to reduce the number of artifacts that could mitigate the identification of important information from the regions that characterize a metastasis and/or an organ at risk. This result was confirmed using a mathematical formalism. Image registration was performed between CT and MRI datasets to align all data from the same patient. As a result, a dataset comprising information on treatments and follow-ups of diverse patients was obtained. This dataset, along with respective transformation matrices, is prepared to be used in various studies related to brain metastases and their evolution over time. RFUds is publicity available at zenodo under the DOI \href{https://zenodo.org/records/14525197}{10.5281/zenodo.14524784}.

\newpage
\bibliographystyle{unsrt}  
\bibliography{references}  

\begin{thebibliography}{10}

\bibitem{catorze}
Metastatic brain tumors | johns hopkins medicine.

\bibitem{sete}
Mark~J. Amsbaugh and Catherine~S. Kim.
\newblock Brain metastasis.
\newblock {\em StatPearls}, 4 2023.

\bibitem{quinze}
Brain metastasis - brainlab.org.

\bibitem{dezasseis}
Imola Wilhelm, Judit Molnár, Csilla Fazakas, János Haskó, and István~A. Krizbai.
\newblock Role of the blood-brain barrier in the formation of brain metastases.
\newblock {\em International Journal of Molecular Sciences}, 14:1383, 1 2013.

\bibitem{dezassete}
R.~Soffietti, A.~Ducati, and R.~Rudà.
\newblock Brain metastases.
\newblock {\em Handbook of Clinical Neurology}, 105:747--755, 1 2012.

\bibitem{dezoito}
Adam Lauko, Yasmeen Rauf, and Manmeet~S. Ahluwalia.
\newblock Medical management of brain metastases.
\newblock {\em Neuro-oncology Advances}, 2, 1 2020.

\bibitem{dezanove}
Xuling Lin and Lisa~M. DeAngelis.
\newblock Treatment of brain metastases.
\newblock {\em Journal of Clinical Oncology}, 33:3475, 10 2015.

\bibitem{vinte}
Alex~W. Brenner and Akash~J. Patel.
\newblock Review of current principles of the diagnosis and management of brain metastases.
\newblock {\em Frontiers in Oncology}, 12, 5 2022.

\bibitem{tres}
Radiation therapy for cancer - nci.

\bibitem{vinteeum}
Radiation therapy | radiation treatment for cancer | american cancer society.

\bibitem{vinteedois}
Types of radiation therapy | seer training.

\bibitem{vinteetres}
Srs and srt: Stereotactic radiosurgery and radiotherapy | stony brook cancer center.

\bibitem{vinteequatro}
Stereotactic radiosurgery - mayo clinic.

\bibitem{vinteecinco}
What is cobalt-60 radiosurgery? - brainlab.org.

\bibitem{vinteeoito}
Types of radiation therapy used to treat metastasis to the brain - brainlab.org.

\bibitem{vinteenove}
What is linear accelerator stereotactic radiosurgery? - brainlab.org.

\bibitem{oitentaecinco}
Dicom in radiotherapy.

\bibitem{oitentaeseis}
Digital imaging and communications in medicine (dicom) supplement 11 radiotherapy objects.

\bibitem{oitentaesete}
Alvaro Rojas-Villabona, Neil Kitchen, and Ian Paddick.
\newblock Investigation of dosimetric differences between the tmr 10 and convolution algorithm for gamma knife stereotactic radiosurgery.
\newblock {\em Journal of Applied Clinical Medical Physics}, 17:217, 2016.

\bibitem{oitentaeoito}
Dicom library - about dicom format.

\bibitem{oitentaenove}
Ps3.15.

\bibitem{noventeeum}
Xiangrui Li, Paul~S. Morgan, John Ashburner, Jolinda Smith, and Christopher Rorden.
\newblock The first step for neuroimaging data analysis: Dicom to nifti conversion.
\newblock {\em Journal of Neuroscience Methods}, 264:47--56, 5 2016.

\bibitem{noventaedois}
Sikerdebaard/dcmrtstruct2nii: v5.

\bibitem{noventaetres}
Tami~D. DenOtter and Johanna Schubert.
\newblock Hounsfield unit.
\newblock {\em Radiopaedia.org}, 3 2023.

\bibitem{noventaequatro}
Yahya Baba and Andrew Murphy.
\newblock Windowing (ct).
\newblock {\em Radiopaedia.org}, 3 2017.

\bibitem{noventaecinco}
Jagriti Kalyani and Monisha Chakraborty.
\newblock Contrast enhancement of mri images using histogram equalization techniques.
\newblock {\em 2020 International Conference on Computer, Electrical and Communication Engineering, ICCECE 2020}, 1 2020.

\bibitem{noventaeseis}
Gabriel Fillipe~Centini Campos, Saulo~Martiello Mastelini, Gabriel~Jonas Aguiar, Rafael~Gomes Mantovani, Leonimer~Flávio de~Melo, and Sylvio Barbon.
\newblock Machine learning hyperparameter selection for contrast limited adaptive histogram equalization.
\newblock {\em Eurasip Journal on Image and Video Processing}, 2019:1--18, 12 2019.

\bibitem{noventa}
Ana~Sofia Santos.
\newblock Predicting brain metastases recurrence after radiation therapy treatment - a processing pipeline.
\newblock 2023.

\bibitem{noventaesete}
A.~Blum, R.~Gillet, A.~Rauch, A.~Urbaneja, H.~Biouichi, G.~Dodin, E.~Germain, C.~Lombard, P.~Jaquet, M.~Louis, L.~Simon, and P.~Gondim Teixeira.
\newblock 3d reconstructions, 4d imaging and postprocessing with ct in musculoskeletal disorders: Past, present and future.
\newblock {\em Diagnostic and Interventional Imaging}, 101:693--705, 11 2020.

\bibitem{noventaeoito}
Min Chen, Nicholas~J. Tustison, Rohit Jena, and James~C. Gee.
\newblock Image registration: Fundamentals and recent advances based on deep learning.
\newblock {\em Neuromethods}, 197:435--458, 7 2023.

\bibitem{noventaenove}
Degrees of freedom - mipav.

\bibitem{cem}
Mark Jenkinson and Stephen Smith.
\newblock A global optimisation method for robust affine registration of brain images.
\newblock {\em Medical Image Analysis}, 5:143--156, 6 2001.

\bibitem{centoeum}
Mark Jenkinson, Peter Bannister, Michael Brady, and Stephen Smith.
\newblock Improved optimization for the robust and accurate linear registration and motion correction of brain images.
\newblock {\em NeuroImage}, 17:825--841, 2002.

\bibitem{centoetres}
Stephen~M. Smith.
\newblock Fast robust automated brain extraction.
\newblock {\em Human brain mapping}, 17:143--155, 11 2002.

\end{thebibliography}

\end{document}